\documentclass[aps,prl,twocolumn,showpacs]{revtex4}
\usepackage{bm,epsfig,amsmath,amssymb}
\newcommand{\half}{\tfrac{1}{2}}

\newcommand{\ez}{{\mathbf {\hat e}_{z}}}

\newcommand{\uvec}{{\mathbf u}}
\newcommand{\xvec}{{\mathbf x}}
\newcommand{\avec}{{\mathbf a}}

\newcommand{\kvec}{{\mathbf k}}

\newcommand{\Uvec}{{\mathbf U}}
\newcommand{\vvec}{{\mathbf v}}

\newcommand{\Nmat}{{\mathcal N}}
\newcommand{\Pmat}{{\mathcal P}}
\newcommand{\Smat}{{\mathcal S}}

\newcommand{\pd}[1]{\frac{\partial{#1}}{\partial t}}
\newcommand{\fd}[1]{\frac{\mathrm{d}{#1}}{\mathrm{d}t}}
\newcommand{\dotp}[2]{  {#1}\cdot {#2}  }
\begin{document}
%
\title{Mean effects of turbulence on elliptic instability in fluids}
\author{Bruce R. Fabijonas$^1$ and Darryl D. Holm$^2$}
\affiliation{
$^1$Department of Mathematics,
Southern Methodist University,
Dallas, TX  \\
$^2$Theoretical Division and Center for Nonlinear Studies,
Los Alamos National Laboratory,
Los Alamos, NM}
\begin{abstract}  
Elliptic
instability in fluids is discussed in the context of the
Lagrangian-averaged Navier-Stokes-alpha (LANS$-\alpha$)
turbulence model.  This model preserves the Craik-Criminale (CC) family of
solutions consisting of a columnar eddy and a Kelvin
wave.
The LANS$-\alpha$ model is shown to preserve the 
elliptic instability for the inviscid case.  However, the model
shifts the critical stability angle.  This shift increases
(resp. decreases) the maximum growth rate for long (resp. short)
waves.  It also introduces a band of stable CC solutions for short waves.
\end{abstract}
\maketitle

Elliptic, or tilting, instability is a fundamental
  phenomenon in fluids  that results from parametric
  resonance. This is the mechanism by which vortex
  stretching creates three dimensional instabilities in
  swirling two dimensional flows. Specifically, the energy in
  an elliptical columnar eddy may be transferred to a
  propagating Kelvin wave \cite{kelvin} by this mechanism. A breakthrough in
  the study of this problem occurred when Craik \&
  Criminale \cite{craik:crim:86} showed  
  that superposing a columnar eddy and a Kelvin wave yields an
  exact solution of the Euler equations. Thus, the elliptical
  instability can  be treated as a modulation of the
  Craik-Crimale (CC) family of solutions by using Floquet
  analysis as was first done by Bayly \cite{bayly:86}.  The history of discovery
  and rediscovery of elliptic instability in laminar fluids is
  reviewed by Kerswell \cite{ker:02}. Here we address the mean
  effects of turbulence on the elliptic instability. The CC
  family of superposed solutions is preserved by the closure
  model we shall consider. For this model, turbulence is shown to
  enhance the growth rates of elliptic instability for Kelvin
  wavelengths that are longer than the turbulence correlation
  length. Conversely, turbulence is found to suppress elliptic
  instability at the shorter wavelengths and to create a band of stable 
  CC flows with nonzero eccentricities.

  The turbulence model we shall consider is the
  Lagrangian-averaged Navier-Stokes-alpha (LANS$-\alpha$)
  model \cite{chen:foias:holm:olson:titi:wynne}, whose equations are:
\begin{multline}
\pd{\vvec} + (\dotp{\uvec}{\nabla})\vvec +
\dotp{(\nabla u)^T}{\vvec}   \\ + 
\nabla \left ( p - \tfrac{1}{2}|\uvec|^2 
- \tfrac{1}{2}\alpha^2|\nabla\uvec|^2 \right ) = \nu\Delta \vvec
 \label{eq:uevol}
\end{multline}
together with $\dotp{\nabla}{\uvec} = 0$.
  In this model, the mean fluid velocity $\mathbf{u}$ is
  related to the mean momentum $\mathbf{v}$ via the Helmholtz
  operator $(1-\alpha^2\Delta)$ as
  $\vvec=\uvec-\alpha^2\Delta\uvec$. This
  Helmholtz filtering of the fluid velocity introduces the
  length scale $\alpha$ as a parameter in the model.
 The LANS$-\alpha$ model preserves the fundamental transport 
 theorems for circulation and vorticity dynamics of the NS 
 equations. Direct numerical simulations of the 
 LANS$-\alpha$ model for forced homogeneous turbulence show
 it to be considerably less computationally intensive 
 than the exact NS equations while preserving essentially the same 
 behavior as NS at length scales larger than alpha \cite{chen:holm:mar:zha:99}. 
 The unforced, inviscid Euler$-\alpha$ form of these equations first 
 appeared in the context of averaged fluid models \cite{holm:mar:rat}. 
 The basic properties of the  LANS$-\alpha$ model, its comparison with
  experiment, and its early 
 development are reviewed in Ref. \onlinecite{foias:holm:titi:01}. See also Refs. 
 \onlinecite{newalpha} 
 for additional results for this model. 
 As discussed in Ref.~\cite{holm:mar:rat} the LANS$-\alpha$ turbulence equations
  formally coincide in the inviscid limit with a classic rheological
  model known as the 2nd-grade fluid \cite{grade2}. Thus, the
  present results for the inviscid elliptic instability apply to both
  the LANS$-\alpha$ turbulence model and to rheology of 2nd-grade
  fluids.

We construct an exact solution to Eq.~\eqref{eq:uevol} with zero divergence  
of the form $\uvec_0(\xvec,t) = \Smat(t)\xvec + \Uvec(t)$, where
$\Smat\xvec$ is the action of the matrix $\Smat$ on the vector 
$\xvec = [x,y,z]^T$ from the left.  
The matrix $\Smat$ is a time dependent matrix with zero trace 
such that $d\Smat/dt +\Smat^2={\mathcal M}(t)$, where 
${\mathcal M}(t)$ is a symmetric matrix which contains the contributions
of $\Uvec(t)$.  
The corresponding pressure $p_0$ is obtained from 
${\mathcal M}(t)$; see Ref. \onlinecite{craik:crim:86} for details.  
We nondimensionalize the system using the variables $\xvec' = \xvec/l$,
$t' = \omega t$, $\uvec' = \uvec/\omega l$, $\vvec' = \vvec/\omega l$,
$\alpha' = \alpha/\sqrt{l}$, where $l$ is a typical length scale and 
$\omega = \tfrac{1}{2}|\nabla\times\uvec_0|$.  The resulting equation with 
the prime notation suppressed is Eq.~\eqref{eq:uevol} with $\nu$ 
replaced by $\nu/\omega$.    
We construct a second solution to Eq.~\eqref{eq:uevol} of the form 
$\uvec_0 + \uvec_1$ with corresponding pressure $p_0 + p_1$, where 
\begin{align}
\uvec_1 &= \mu \avec(t)\sin(\beta\psi(\xvec,t)), \\ 
\begin{split}
        p_1 &= \mu \hat p_{11}(t)\cos(\beta\psi(\xvec,t)) \\
        &\phantom{XXXXXXX}+ \mu^2\hat p_{12}(t)\cos^2(\beta\psi(\xvec,t)),
\end{split}
\end{align}
$\psi(\xvec,t) = \dotp{\kvec(t)}{\xvec} + g(t)$, and 
$\mu$ and $\beta$ are scaling factors so that we can choose the initial 
conditions $|\avec(0)| = 1$ and $|\kvec(0)| = 1$.  
The unknown phase $\psi(\xvec,t)$ and the amplitudes
$\avec(t)$, $\hat p_{11}(t)$, and $\hat p_{12}(t)$ are to be determined.
Such flows which are the 
sum of a `base flow' $\uvec_0$ and a `disturbance' $\uvec_1$ 
are called $\alpha$-CC flows.  The incompressibility condition gives 
\begin{eqnarray}
\dotp{\avec}{\kvec} = 0 \label{eq:incomp}.
\end{eqnarray}
The evolution equations for the amplitudes and phase are 
\begin{align}
(\partial_t + \dotp{\Smat\xvec}{\nabla} )\psi + \dotp{\Uvec}{\kvec}&= 0 \label{eq:xievol}, \\
\begin{split}
(\partial_t + \dotp{\Smat\xvec}{\nabla} )((1+\Gamma)\avec) 
        + \Gamma&\Smat^T\avec 
         + \Smat \avec \\
        - (\beta \hat p_{11}
        - \beta^2\alpha^2\dotp{\avec}{\Smat\kvec})\kvec &= 
        - \frac{(1+\Gamma)\nu\beta^2}{\omega}|\kvec|^2\avec ,
\end{split}
 \label{eq:preaevol} \\
\hat p_{12} - \Gamma|\avec|^2 & = 0.\label{eq:p2evol} 
\end{align}
Here $\Gamma = \alpha^2\beta^2|\kvec|^2$.  
Note that the amplitude scaling $\mu$ is 
immaterial.  The parameter $\alpha$ couples various terms throughout the 
system. Moreover, this coupling in $\alpha$ appears only in the 
combination $\Gamma$, which is proportional to 
wavenumber-squared. Consequently, this coupling   
affects the high wavenumber behavior of the solution for $\alpha\ne 0$.
Equation \eqref{eq:xievol} states that the phase is advected with the 
base flow.
Only two free parameters remain 
($\Gamma$ and $E_\omega$) upon introducing 
the vorticity based Ekman number 
$E_\omega = \nu\beta^2/\omega$.
Without loss of generality, we set 
$\partial g/\partial t + \dotp{\kvec}{\Uvec} = 0$.
Denoting the material
derivative as ${\mathrm d_t} = (\partial_t + \dotp{\Smat\xvec}{\nabla})$
and taking the gradient of Eq.~\eqref{eq:xievol} reduces 
Eqs.~\eqref{eq:xievol}-\eqref{eq:preaevol} 
to a system of ordinary differential equations:
\begin{align}
\fd{\kvec} + \Smat^T \kvec &= 0, \label{eq:kevol}\\
\begin{split}
\fd{((1+\Gamma)\avec)} + (1+\Gamma)\Smat^T \avec &\\ +  2\varpi\times\avec 
        - \tilde P\kvec &= 
        - (1+\Gamma) E_\omega|\kvec|^2\avec, \label{eq:aevol}
\end{split}
\end{align}
where $\tilde P$ is the coefficient of $\kvec$ in Eq.~\eqref{eq:preaevol},
$\varpi = \tfrac{1}{2}\nabla\times\uvec_0$ is the (normalized) vorticity of
the base flow and $(\Smat - \Smat^T)\avec = 2\varpi\times\avec$.
We eliminate the pressure term by 
taking the dot product of Eq.~\eqref{eq:aevol} with $\kvec$ and by using 
 $\dotp{{\mathrm d}\avec/{\mathrm dt}}{\kvec} = - \dotp{\avec}{
{\mathrm d}\kvec/{\mathrm dt}} =\dotp{\Smat\avec}{\kvec}$, the 
first of which follows from Eq.~\eqref{eq:incomp} and the second from 
Eq.~\eqref{eq:kevol}: 
\begin{eqnarray}\label{eq:pressure}
\tilde P = 
        \frac{1}{|\kvec|^2}\Big \{ (1+\Gamma)\dotp{(\Smat+\Smat^T)\avec}{\kvec}
        + 2\dotp{\varpi\times\avec}{\kvec} \Big \}.
\end{eqnarray}
In summary, we have obtained a new exact incompressible solution to Eq.~\eqref{eq:uevol}.
The variables are amplitude $\avec(t)$ and wave vector $\kvec(t)$.  Once 
these are determined, the pressure terms follow from Eqs.~\eqref{eq:p2evol} and 
\eqref{eq:pressure}.  
Note that $\uvec_0$ and $\uvec_0 + \uvec_1$ 
are exact solutions to the nonlinear equations,
but $\uvec_1$ by itself is only a solution to Eq.~\eqref{eq:uevol} linearized 
about $\uvec_0$.  The construction described above also can be applied to 
Eq.~\eqref{eq:uevol} expressed in a rotating coordinate system in which 
an $\alpha$-CC flow can still be found.  The effects of rotation will
be discussed elsewhere. Finally, we emphasize that the operator
${\mathrm d}/{\mathrm dt} + S^T$ acting on a vector represents the
complete time 
derivative of that quantity in a Galilean frame moving with $\uvec_0$.

Insight into the dynamics of the problem can be gained by examining
Eq.~\eqref{eq:aevol} in the asymptotic
regimes $\Gamma \ll 1$ and $\Gamma \gg 1$, where  
 $\Gamma = \alpha^2\beta^2|\kvec|^2$.  
(We assume that $|\kvec(t)|$ remains
bounded and never vanishes.)  
For $\Gamma \ll 1$,  Eq.~\eqref{eq:aevol} becomes 
\begin{multline}\label{eq:smallab}
\fd{\avec}  +\Smat^T\avec = -E_\omega|\kvec|^2\avec 
 - 2\varpi\times\avec+ \frac{2\dotp{\Smat\avec}{\kvec}}{|\kvec|^2}\kvec \\
+ \Gamma \left
( 2\varpi\times\avec +
\frac{2}{|\kvec|^2}\left \{ (\dotp{\Smat\kvec}{\kvec})\avec 
   - (\dotp{\varpi\times\avec}{\kvec})\kvec
\right \}  \right ) \\+ O\left ( \Gamma^2 \right ).
\end{multline}
Combined with Eq.~\eqref{eq:kevol}, 
this equation preserves $\dotp{\avec}{\kvec}=0$ at each order.  
The $O(1)$ term
in the above equation is exactly the expression for the amplitude of the 
modulated traveling wave in the CC flow for the classical Euler and
NS equations.  This, of course, is expected since Eq.~\eqref{eq:uevol} reduces
to the NS equations for $\alpha= 0$.  
Thus, to leading order,  the amplitude 
decays with viscosity,
stretches with the base shear and rigidly rotates with 
the vorticity of the base flow.
For $\Gamma \gg 1$, Eq.~\eqref{eq:aevol} becomes
\begin{multline}\label{eq:bigab}
\fd{\avec} +\Smat^T\avec  = -E_\omega|\kvec|^2\avec 
+ \frac{2\dotp{\Smat\avec}{\kvec}}{|\kvec|^2}\kvec 
  \\ + \frac{2}{|\kvec|^2}\left \{
(\dotp{\Smat\kvec}{\kvec})\avec
-(\dotp{\varpi\times\avec}{\kvec}) \kvec \right \} + O\left (
\frac{1}{\Gamma} \right ) .
\end{multline}
Again, this equation preserves $\dotp{\avec}{\kvec}=0$ at each order
Thus, as $\Gamma\to\infty$ (corresponding to either $\alpha\to\infty$
or $\beta\to\infty$), the amplitude no longer rigidly rotates with 
the vorticity of the base flow.

As an example, we examine the stability of a rotating column of fluid with 
elliptic streamlines whose foci lie on the $y$-axis and vorticity
$\varpi = \omega\ez$:  
\begin{eqnarray}
\uvec_0 = \omega L\xvec, \quad L = \begin{pmatrix} 
 0 & -1+\gamma & 0 \\ 1+\gamma & 0 & 0 \\
        0 & 0 & 0 \end{pmatrix}.
\end{eqnarray}
Here, $0 \leq \gamma < 1$ is the eccentricity of the ellipses, and the pressure
is $p_0 = \tfrac{1}{2}\omega^2 (1-\gamma^2)(x^2+y^2)$. 
Equation \eqref{eq:kevol} with $\Smat=L$ is analytically solvable:
\begin{multline} \label{eq:ksol}
\kvec = [\sin\theta\cos(t\sqrt{1-\gamma^2}), \\
        \kappa\sin\theta\sin(t\sqrt{1-\gamma^2}), \cos\theta]^T 
\end{multline}
where $\kappa^2 = (1-\gamma)/(1+\gamma)$ and $\theta$ is 
the polar angle that $\kvec$ makes with the 
axis of rotation.  In summary, we have a four parameter problem in 
$\Gamma$, $E_\omega$, $\gamma$, and $\theta$.   
Eq.~\eqref{eq:aevol} has the 
form ${\mathrm d}\avec/{\mathrm dt} = \Nmat(t) \avec$, where the 
elements of the matrix $\Nmat(t)$ are periodic with period 
$\tau = 2\pi/\sqrt{1-\gamma^2}$.  Therefore, the system can be analyzed 
numerically using Floquet theory \cite{yaku:star:76}.  We compute the 
monodromy matrix $\Pmat$, that is, the fundamental solution matrix
with identity initial condition evaluated at $t = \tau$.  Equation
\eqref{eq:aevol} will have exponentially growing solutions if 
$\max_i|\Re (\rho_i)| > 1$, where $\rho_i, i=1,2,3$ are the eigenvalues
of $\Pmat$, with corresponding Lyapunov-like growth rates given by 
\begin{eqnarray}
\sigma = \frac{1}{\tau}\ln \{\max_i|\Re (\rho_i)|\}.
\end{eqnarray}
Thus, we can simulate numerically the solution to Eq.~\eqref{eq:aevol} 
over one period and indisputably determine the exponential 
growth rates.  
We can be certain that at least one of the eigenvalues will always 
be unity because 
of the incompressibility condition \eqref{eq:incomp} and 
that the remaining two eigenvalues 
appear as complex conjugates on the unit circle or as real valued 
reciprocals of each other.  

The present investigation 
considers the case of inviscid flow, i.e. $E_\omega = 0$.
Viscosity, which only  slightly modifies the inviscid results, will be
discussed elsewhere.
For flows with circular streamlines ($\gamma = 0$), 
the monodromy matrix can be analytically computed. 
It follows from Eq.~\eqref{eq:ksol} 
that $|\kvec(t)| = 1$.
Then, $\Gamma$ 
is constant in time (denoted by $\Gamma_0 = \alpha^2\beta^2$) and
Eq.~\eqref{eq:aevol} has   
three linearly independent solutions:
\begin{align}
\avec_1(t) &= \cos(\xi(t)+\phi) \kvec_{\perp 1} + \sin(\xi(t)+\phi)
\kvec_{\perp 2} \label{eq:a1} \\
\avec_2(t) &= \sin(\xi(t)+\phi) \kvec_{\perp 1} - \cos(\xi(t)+\phi) \kvec_{\perp 2} \\
\avec_3(t) &= \ez,
\end{align}
where $\xi(t) = 2t\cos\theta /(1+\Gamma_0)$, 
$\kvec_{\perp 2} = [\sin t, -\cos t, 0]^T$ and  
$\kvec_{\perp 1} = [\cos\theta \cos t, \cos\theta \sin t,
-\sin\theta]^T$ are vectors orthogonal to $\kvec$, and $\phi$ is an arbitrary phase. 
Clearly the first two solutions $\avec_1$ and $\avec_2$
satisfy Eq.~\eqref{eq:incomp}.  The monodromy matrix can be constructed from
these three solutions:
\begin{eqnarray*}
\Pmat = \begin{pmatrix}
\cos(\xi(2\pi)) & \cos\theta\sin(\xi(2\pi)) & 0 \\
-\sin(\xi(2\pi))/\cos\theta & \cos(\xi(2\pi)) & 0 \\
\tan\theta (1-\cos(\xi(2\pi))) & -\sin\theta\sin(\xi(2\pi)) & 1 
\end{pmatrix}.
\end{eqnarray*}
The three eigenvalues are 
$\rho_{1,2} = \exp(\pm i\xi(2\pi)), \rho_3 = 1$.  
All of the eigenvalues lie on the unit circle, 
from which it follows that all solutions 
in the inviscid case for $\gamma = 0$ are stable.  The values 
of $\cos\theta$ for which $|\rho_i| = 1, i=1,2,3$ are called `critically
stable' and are given by $\xi(2\pi) = n\pi$, $n = 0, \pm 1, \pm 2, \ldots$, 
corresponding to 
$\cos\theta = n(1+\Gamma_0)/4$.  
At these parameter values an exponentially growing solution can 
appear (together with an exponentially decaying one) 
as $\gamma$ increases from zero.
Since $\Gamma_0 \geq 0$, the only
values of interest are $n = 0, \pm 1, \pm 2, \pm 3$, 
and, for the case $\alpha = 0$, $n= \pm 4$.
Bayly \cite{bayly:86} argues that the evenness of 
$\tilde P\kvec$ as a function of $\kvec$ implies that the
eigenvalues, if real and unequal, must be positive.  This dismisses 
the $n = \pm 1$ and $n=\pm 3$ choices.  
The cases $n=0$ and $n = \pm 4$ preserve the two-dimensional structure
of the base flow and thus should be stable under small 
perturbations in the eccentricity.
The remaining value, $\cos\theta = \tfrac{1}{2}(1+\Gamma_0)$ is 
the critical parameter value at which $\avec(t)$ suffers exponential
growth as $\gamma$ increases from zero.  
We conclude that introducing $\alpha$ preserves the existence of 
elliptic instability, though
it shifts the angles at which elliptic instability 
arises to $\cos\theta =
(1+\Gamma_0)/2$.  
In addition, for $\Gamma_0 > 1$, the LANS$-\alpha$
model stabilizes 
Bayly's elliptic instability. 

\begin{figure}[ht]
\begin{center}
\epsfig{file=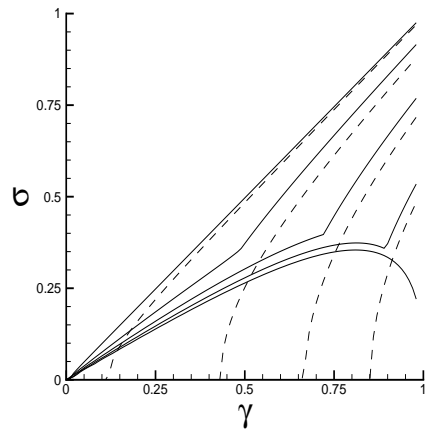,width=2.5in,height=2.5in} 
\caption{The growth rate maximized over $\cos\theta$ for $E_\omega =
0$ and several
values of $\Gamma_0 = \alpha^2\beta^2$.  The solid lines are, from bottom to top,
$\Gamma_0 = 0, 0.1, 0.25, 0.5, 1$.  The maximum growth rate is bounded by
Eq.~\eqref{eq:grrate}. The dashed lines, from top to bottom, are
$\Gamma_0 = 1.25, 2.5, 5.0, 12.5$.  Notice that for $\Gamma_0 > 1$, a
stable band of nonzero eccentricities appears.  
} 
\label{fig:growth}
\end{center}
\end{figure}

Additional understanding of this result emerges by following the
analysis of 
Waleffe \cite{wale:90} and Kerswell \cite{ker:02}.  
By taking the dot product of Eq.~\eqref{eq:aevol} with $\avec$, we
obtain (for all $\gamma$)
\begin{eqnarray}
\fd{\left ( \half |\avec|^2 \right ) } = -2\gamma a_1 a_2 +
\frac{4\gamma\Gamma}{1+\Gamma}\frac{k_1k_2}{|\kvec|^2}|\avec|^2. 
\end{eqnarray}
One can determine an exponential growth rate to leading order in
$\gamma$ by inserting the zeroth order solutions for $\kvec$ and
$\avec_1$ into the right hand side of this equation:
\begin{multline}
\sigma \equiv \frac{1}{|\avec|^2}\fd{\left ( \half|\avec|^2\right ) }
 = -\frac{\gamma}{4}[(1-\cos\theta)^2\sin(2(\xi_++\phi)) \\
   - (1+\cos\theta)^2\sin(2(\xi_-+\phi)) \\
   - 2(1-\cos^2\theta)\sin(2t)]
  + \frac{2\gamma\Gamma_0}{1+\Gamma_0}\sin^2\theta\sin(2t),
\end{multline}
where $\xi_\pm = \xi(t) \pm t$.  Upon averaging over a period of
$\avec_1$, this quantity will vanish 
except when $\xi_\pm = 0$, corresponding to $\cos\theta =
\mp\Gamma_0/2$.  The 
maximum values for $\sigma$ will occur at $\phi = \mp\pi/4$ for
$\xi_\pm = 0$, respectively, with growth rate
\begin{eqnarray}\label{eq:grrate}
\sigma = \frac{(3+\Gamma_0)^2}{16}\gamma + O\Big(\gamma^2\Big ),
\end{eqnarray}
valid for $\Gamma_0 \leq 1$.  Thus, we see that the angle of critical
stability is again $\cos\theta = \mp(1+\Gamma_0)/2$.  Furthermore, we
see that the maximum growth rate increases as a function of
$\Gamma_0$ due to the $\Gamma_0$ dependence of the critical stability point
up to a maximum of $\sigma = \gamma$ at $\Gamma_0 = 1$,
after which a set of stable solutions emerges in a band of nonzero
eccentricities.  See Fig.~\ref{fig:growth}. 

\begin{figure}[ht]
\begin{center}
\epsfig{file=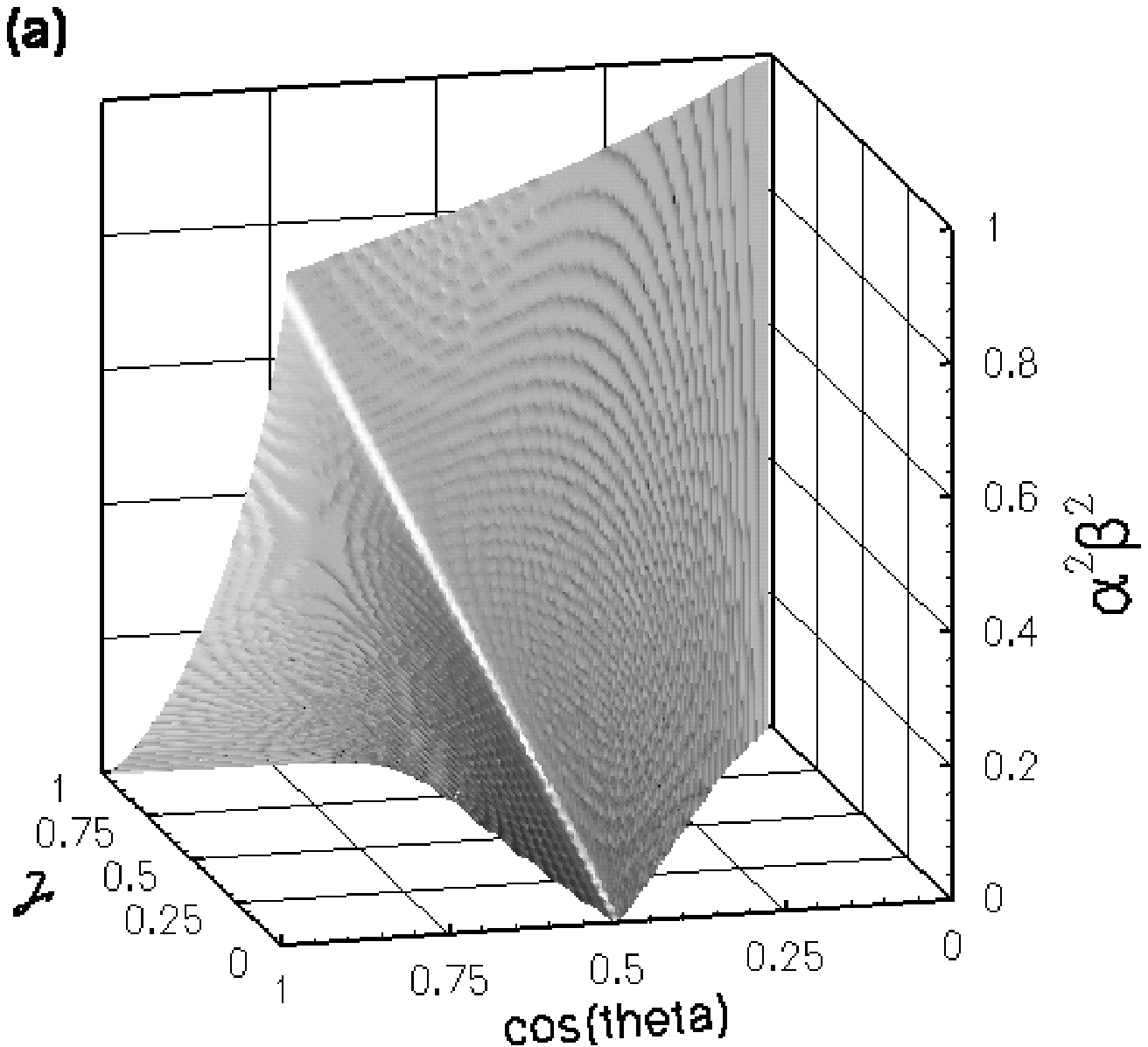,width=2.5in,height=2.5in}
\epsfig{file=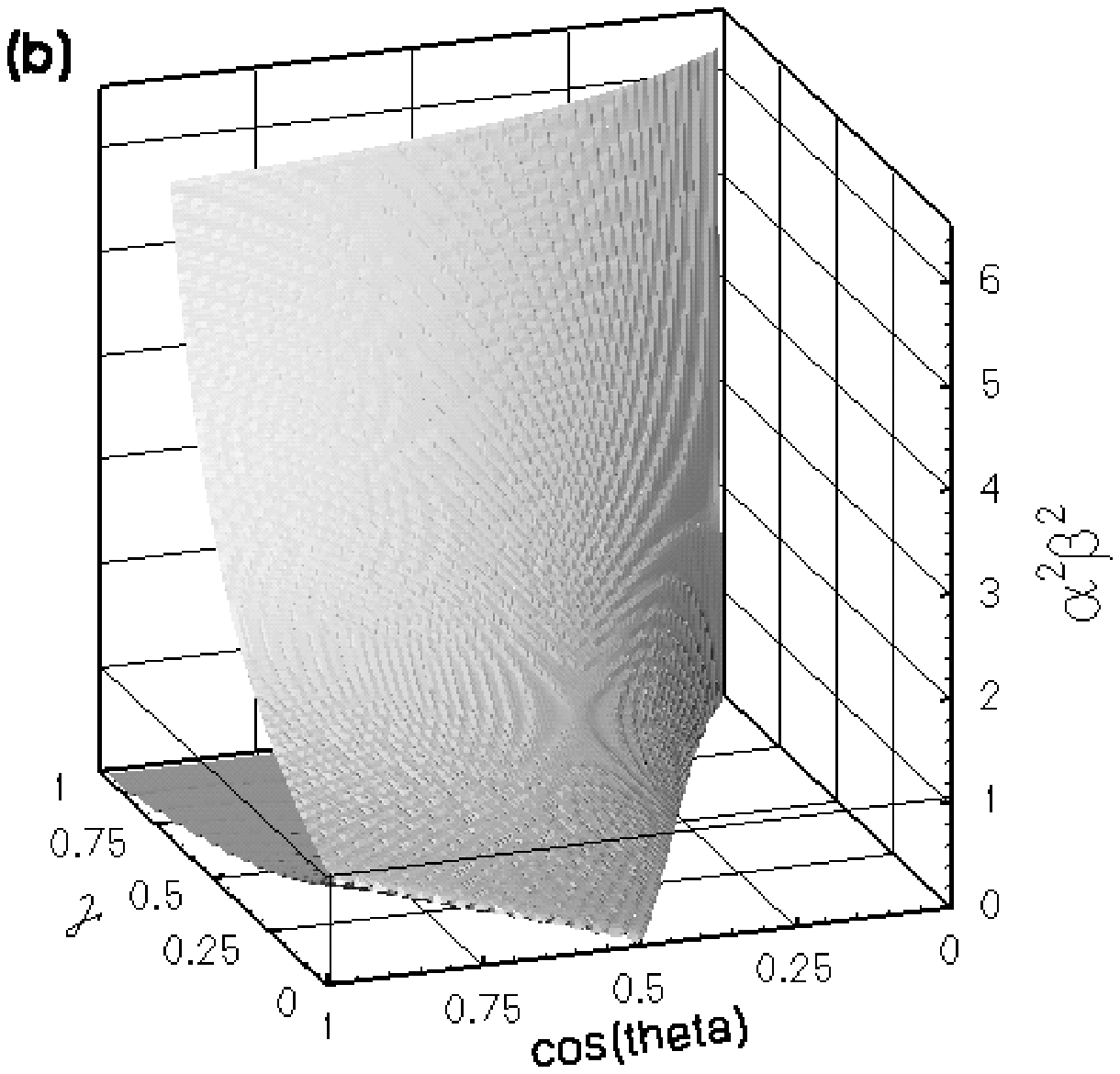,width=2.5in,height=2.5in} 
\caption{Surface of $\sigma = 0.01$ for $E_\omega = 0$.  
The horizontal plane is the $\gamma$-$\cos\theta$
plane and the vertical axis is $\alpha^2\beta^2$.  
Figure (a) shows the neutral surface for $0 \leq \alpha^2\beta^2 \leq 1$ and
is an expansion of the boxed region in figure (b).  
For $\alpha = 0$, the critical stability point 
occurs at $\theta = \pi/3$, which agrees with the classical results.  
The critical stability point shifts towards $\cos\theta = 1$
as $\alpha^2\beta^2$
increases according to $\cos\theta = (1+\alpha^2\beta^2)/2$.  As $\alpha^2\beta^2$ exceeds
unity, a stable band of rotating flows with nonzero
eccentricities appears.   
} 
\label{fig:bigfigs1}
\end{center}
\end{figure}

For nonzero values of $\gamma$, we must investigate the system
numerically.  
Figure \ref{fig:bigfigs1} shows the evolution 
of the critical instability surface as a function of $\alpha^2\beta^2$.
For $\alpha^2\beta^2 << 1$, there is little
change in the critical instability surface as predicted by
Eq.~\eqref{eq:smallab}.  For $\alpha^2\beta^2 > 0$, all angles of
incidence for the traveling wave are unstable in a neighborhood of $\gamma = 1$.
The maximum growth 
rate in the $\gamma$-$\cos\theta$ plane increases as a function of
$\alpha^2\beta^2$ and shifts to the corner $\gamma =1$, $\cos\theta =
1$ by $\alpha^2\beta^2=0.1$.   When $\alpha^2\beta^2$ exceeds unity,
the flow stabilizes.  For a given set 
of parameters $(\gamma,\cos\theta)$, one of following three situations 
will occur as shown in Fig.~\ref{fig:bigfigs1}:
the flow is stable for all $\alpha^2\beta^2$; the flow is unstable for  $0 \leq
\alpha^2\beta^2 < \alpha_1^2\beta_1^2$ and stable for
$\alpha^2\beta^2\geq\alpha_1^2\beta_1^2$; or the flow is 
stable for $0\leq\alpha^2\beta^2\leq\alpha_2^2\beta_1^2$, unstable for
$\alpha_2^2\beta_2^2<\alpha^2\beta^2<\alpha_1^2\beta_1^2$, and stable again for
$\alpha^2\beta^2\geq\alpha_1^2\beta_1^2$.

Thus, the LANS$-\alpha$ turbulence model enhances the growth rates of
the elliptic instability for long waves with $\alpha^2\beta^2 < 1$
while it shifts the angle of critical stability along the cusp rising
diagonally in 
Fig.~\ref{fig:bigfigs1}.  It also stabilizes the elliptic instability
for short waves with $\alpha^2\beta^2 > 1$ as seen in
Fig.~\ref{fig:bigfigs1}b.  Finally, for any $\alpha^2\beta^2 \neq 0$, this
turbulence model modifies the region in $(\gamma,\cos\theta)$ parameter space where the
elliptic instability occurs, as also shown in Fig.~\ref{fig:bigfigs1}.

In principle, one can now examine the stability properties of this new
family of exact $\alpha-$CC
solutions of the LANS$-\alpha$ model.  
This would be a secondary stability analysis of the rotating base flow.
Work of this type was carried out by Lifschitz and 
collaborators \cite{lif:fab}
for the classical CC flows under
high-frequency, short wavelength perturbations.
A similar perturbation analysis for the $\alpha$-CC flow
will be carried out elsewhere. 

The authors are indebted to A. Lifschitz-Lipton for stimulating our
original interest in CC solutions
and to the Center for Scientific Computing
at Southern Methodist University for use of their computing
resources.  Furthermore, BF thanks the Theoretical Division at the 
Los Alamos National 
Laboratory for their hospitality.  The solutions to $\avec(t)$ 
were
simulated using the variable coefficient ODE solver \textsf{DVODE} \cite{dvode}.

\end{document}